\documentclass[sigconf]{acmart}

\usepackage{amsmath}

\usepackage{algorithm}
\usepackage{algorithmicx}
\usepackage{algpseudocode}

\usepackage{graphicx}
\usepackage{textcomp}

\usepackage{booktabs} 
\usepackage{listings}
\usepackage{subcaption}
\usepackage{multirow}
\usepackage{balance}
\usepackage{url}
\usepackage[multiple]{footmisc}
\usepackage[color=white]{todonotes}
\usepackage[export]{adjustbox}
\usepackage{xspace}
\usepackage{tcolorbox}
\usepackage[acronym, nopostdot, nonumberlist, nogroupskip, section=section]{glossaries}
\usepackage{array}
\usepackage{wrapfig}
\usepackage{tabu}
\usepackage{enumitem}
\usepackage{silence}
\usepackage{booktabs, multirow} 
\usepackage{soul}
\usepackage{changepage,threeparttable} 
\usepackage{listings}
\usepackage{color}
\usepackage{titletoc}
\usepackage{lipsum}
\usepackage{comment}
\usepackage{chngcntr}

\usepackage{tabularx}

\titlecontents{chapter}
  [0pt]
  {}
  {\bfseries\chaptername\ \thecontentslabel:\quad}
  {}
  {\normalsize\titlerule*[10pt]{.}\contentspage}

\makeatletter
\newcommand*{\rom}[1]{\expandafter\@slowromancap\romannumeral #1@}
\makeatother

\lstdefinelanguage{yaml}{
  keywords={true,false,null,y,n},
  keywordstyle=\color{blue},
  basicstyle=\ttfamily\small,
  sensitive=false,
  comment=[l]{\#},
  morecomment=[s]{/*}{*/},
  commentstyle=\color{gray},
  stringstyle=\color{red},
  showstringspaces=false,
  identifierstyle=\color{black},
  morestring=[b]',
  morestring=[b]"
}

\definecolor{dkgreen}{rgb}{0,0.6,0}
\definecolor{gray}{rgb}{0.5,0.5,0.5}
\definecolor{mauve}{rgb}{0.58,0,0.82}

\begin{document}

\fancyhead{}

\title{Employing Software Diversity in Cloud Microservices to Engineer Reliable and Performant Systems}

\author{Nazanin Akhtarian}
\affiliation{Department of Computer Science \\  
York University, Toronto, Canada}
\email{nakhtari@yorku.ca}

\author{Hamzeh Khazaei}
\affiliation{Department of Computer Science \\  
York University, Toronto, Canada}
\email{hkh@yorku.ca}

\author{Marin Litoiu}
\affiliation{Department of Computer Science \\  
York University, Toronto, Canada}
\email{mlitoiu@yorku.ca}

\begin{abstract}
In the ever-shifting landscape of software engineering, we recognize the need for adaptation and evolution to maintain system dependability. As each software iteration potentially introduces new challenges, from unforeseen bugs to performance anomalies, it becomes paramount to understand and address these intricacies to ensure robust system operations during the lifetime. This work proposes employing software diversity to enhance system reliability and performance simultaneously.
A cornerstone of our work is the derivation of a reliability metric. This metric encapsulates the reliability and performance of each software version under adverse conditions. Using the calculated reliability score, we implemented a dynamic controller responsible for adjusting the population of each software version. The goal is to maintain a higher replica count for more reliable versions while preserving the diversity of versions as much as possible. This balance is crucial for ensuring not only the reliability but also the performance of the system against a spectrum of potential failures. In addition, we designed and implemented a diversity-aware autoscaling algorithm that maintains the reliability and performance of the system at the same time and at any scale.  Our extensive experiments on realistic cloud microservice-based applications show the effectiveness of the proposed approach in this paper in promoting both reliability and performance.
\end{abstract}



\maketitle

\section{Introduction} \label{introduction}
In the contemporary software landscape, microservices have emerged as a preferred architectural style, enabling modularity, scalability, and easy maintainability. These fine-grained services encapsulate specific functionalities, making systems more flexible and manageable. As the adoption of microservices grew, so did the need for efficient orchestration tools to manage these modular components, especially when housed within containers. Kubernetes\footnote{\href{https://kubernetes.io}{https://kubernetes.io}} is a leading orchestration platform designed specifically for container management, ensuring smooth deployment, scaling, and operation of applications built using microservices.

Beyond architecture and orchestration, the appreciation for multi-version software systems is on the rise. This approach, involving the concurrent maintenance of multiple software versions, enables systems to intelligently select and deploy the most suitable software version based on real-time conditions and metrics, addressing challenges related to performance optimization, fault tolerance, reliability, security vulnerabilities, and continuous availability \cite{b24, b22, b23, b13, b18}. Central to this strategy is the challenge of preserving system reliability, especially given the intricate layers added by multiple software versions. This work introduces several novel contributions to enhance system reliability and performance:

\begin{itemize}
    \item \textbf{Conceptualization of Multi-version Containers:} Introducing transparency in multi-versioning at the container level, allowing users to interact with services without concern for underlying version differences.

    \item \textbf{Dynamic Controller Implementation:} A controller to adjust the population of each software version based on its reliability score, aiming to maintain a higher replica count for more reliable versions while preserving version diversity.
    
    \item \textbf{Diversity Emphasis:} Introduction of the Diversity Factor (DF), which quantifies the distribution of software versions, advocating for a balanced deployment approach over a solely reliability-centric one.
 
    \item \textbf{Diversity-Aware Auto-Scaling Algorithm:}
    Drawing inspiration from natural selection, this framework dynamically allocates resources to the most reliable deployments, mirroring the survival of the fittest in natural ecosystems. Our diversity-aware autoscaling algorithm is designed to maintain system reliability and performance simultaneously, regardless of scale.

\end{itemize}

This work proposes employing software diversity to enhance system reliability and performance simultaneously. Using the calculated reliability score, we implemented a dynamic controller responsible for adjusting the population of each software version. The goal is to maintain a higher replica count for more reliable versions while preserving the diversity of versions as much as possible. This balance is crucial for ensuring the reliability of the system against a spectrum of potential failures. Our extensive experiments on realistic cloud microservice-based applications show the effectiveness of our proposed approach. All the artifacts, including source codes and documentation, related to this paper are publically available to facilitate the reproducibility of our proposed methodology\footnote{https://github.com/nazanin97/ReplicaBalancer/tree/main}.

The rest of the paper is organized as follows. Section~\ref{background} provides background information about multi-version software systems, microservices, auto-scaling, and load balancing. In Section~\ref{approach}, we present the concept of our approach. In Section~\ref{setup}, we explain our experimental setup. Section~\ref{results} discusses the results of our experiments. Section~\ref{related} gives an overview of the related work, and Section~\ref{disc} explains the threats to validity and the future work. Finally, Section~\ref{conclusion} concludes the paper.

\section{Background} \label{background}
This section provides the central concepts in our study and establishes the link between them and software diversity.

\subsection{Microservice Architecture}
Microservices have become an essential paradigm in software design and architecture. Microservices architecture breaks down a complex system into smaller and independent services. Each microservice has its own functionality and can be developed, deployed, and scaled independently, allowing for better resource utilization and cost efficiency \cite{Balalaie}. This approach allows for greater flexibility, scalability, and maintainability compared to traditional monolithic architectures. In addition, microservices can be developed using different technologies and programming languages, allowing teams to choose the most suitable tools for each service. Given this structure, software multi-versioning can be more selectively applied to individual components rather than the system as a whole. So, in this work, we choose our subject system an application with microservice architecture.

\subsection{Auto-Scaling}
In the context of software multi-versioning and microservices, understanding the role of auto-scaling strategies is vital for performance and resource optimization. Elasticity in cloud computing addresses unpredictable workloads, allowing for the adjustment of resources in tune with workloads to balance service performance and costs \cite{fernandez, cruz}. This dynamic adjustment is crucial not only for meeting demand but also for selecting and scaling from the most reliable and efficient versions of services, thereby enhancing the overall stability and performance of the system. In environments where workloads fluctuate, the capability to dynamically scale resources, both at the microservices level and across different software versions, becomes essential for maintaining service stability.

The importance of auto-scaling algorithms in cloud computing is increasing. Proper provisioning of replicas is crucial: under-provisioning can deteriorate performance and risk service unavailability, causing revenue losses \cite{hybrid}, while over-provisioning leads to resource waste and higher costs. Therefore, auto-scaling is essential in optimizing Pod replica resources and avoiding service level objective (SLO) violations.

Auto-scaling can be reactive or proactive \cite{deeplearning}. \textbf{Reactive techniques} analyze the system's current status and decide the scaling based on predefined rules or thresholds. \textbf{Proactive techniques} examine the historical data, predict the future, and perform scaling decisions in advance. This work primarily employs reactive auto-scaling to balance efficiency and simplicity.

\subsection{Load Balancing} \label{sec: Load Balancing}
Load balancing distributes incoming network traffic across multiple servers to ensure no single server is overwhelmed. This maintains the availability and reliability of applications and services. Common methods include Round Robin, Least Connections, Least Response Time, IP Hash, and URL Hash \cite{elgili2020}. Adaptive methods like Weighted Round Robin or Weighted Least Connections dynamically distribute traffic based on server performance or workload fluctuations \cite{harjanti2022load}.
In this work, we use adaptive load balancing with a Weighted Round Robin approach to distribute load based on the reliability measure of underlying software versions.

In our work, we adopt adaptive load balancing, specifically using a Weighted round-robin for request distribution. The load balancer in our approach distributes the load based on the reliability measures of underlying software versions.

\section{Our Approach} \label{approach}
This work introduces an evolutionary strategy inspired by natural selection principles for Kubernetes Pod scaling. This scheme rewards deployments with higher reliability metrics with additional Pods. Central to our solution is a self-adaptive framework encompassing the MAPE (Monitor, Analyze, Plan, Execute) loop:

\begin{itemize}
    \item Monitor: Using Prometheus, we continuously monitor metrics like Pod restarts and memory consumption. Nginx logs are used for client-side response times, and system workloads are observed for scaling insights.
    \item Analyze: We assess Kubernetes deployments' reliability based on a weighted average of three metrics, evaluating workload for scaling needs.
    \item Plan: Based on reliability scores, we decide on adjusting Pod replica counts or traffic distribution among software versions.
    \item Execute: Changes are made to replica counts and load balancer configurations.
\end{itemize}

\subsection{System Architecture} \label{sec:architecture}
\begin{figure}[htbp]
\includegraphics[width=0.48\textwidth]{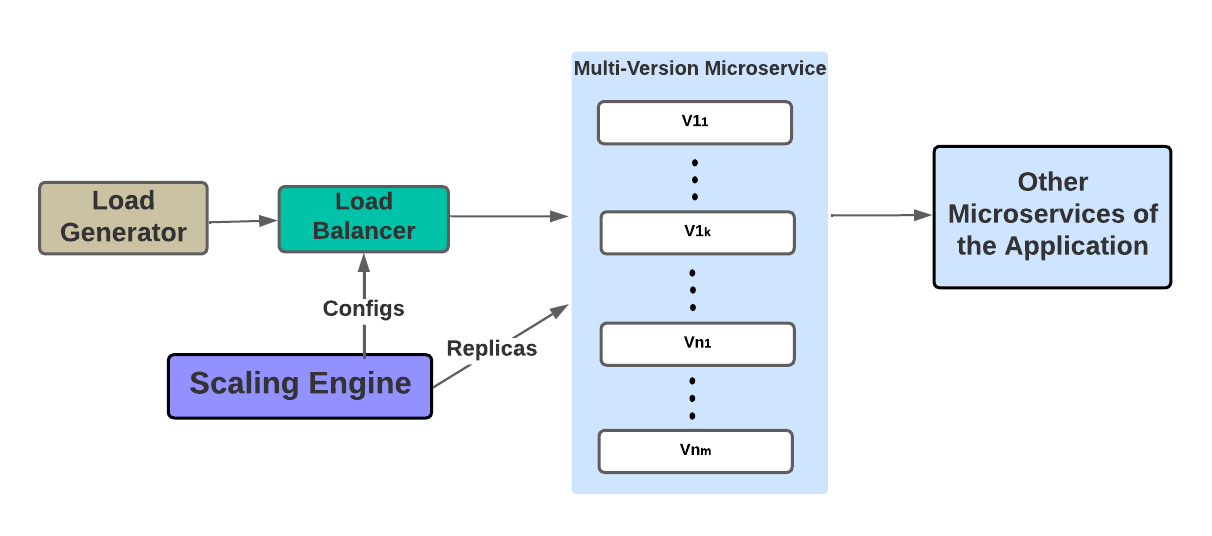}
\caption{System architecture. The diagram illustrates the structure of the proposed solution, detailing component interactions and data flow paths.}
\label{fig:system_architecture}
\Description{System architecture}
\end{figure}

Our solution has been materialized in two fundamental components: the \textbf{Load Balancer} and the \textbf{Scaling Engine}, depicted in Figure \ref{fig:system_architecture}. The architecture distributes the workload dynamically based on microservice version reliability scores and manages replicas accordingly. The load generator drives the traffic to an Nginx-powered\footnote{\href{https://www.nginx.com}{https://www.nginx.com}} load balancer, which uniformly divides requests across multi-version backend servers. The Scaling Engine calculates the reliability score for microservice versions and adjusts replicas based on this score. It also continually assesses and adjusts the overall Pod replica count based on workload monitoring, enhancing system resilience to variable demands and optimizing the allocation of resources. Additionally, it provides configurations to Nginx for weight distribution.

\subsection{Auto-scaling Engine}
Our scaling engine has two core functionalities:
\begin{enumerate}
    \item It dynamically updates the load balancer configurations based on deployment reliability scores to ensure balanced traffic distribution across Kubernetes Pods.
    \item It employs the monitoring system to construct a real-time reliability scoring mechanism based on three metrics: restart counts, response time variability, and memory usage.
\end{enumerate}

\subsubsection{\textbf{Continuous Configuration of Load Balancer at Runtime}} \label{sec: ConfigurationUpdate}
The system proactively updates the load balancer settings to capture the reliability scores continuously. This dynamic adaptation ensures deployments with superior reliability scores handle a larger share of traffic, thereby improving overall system performance.

\subsubsection{\textbf{Reliability Scoring System}}
We devised a system to evaluate the reliability of Kubernetes deployments methodically, calculating a weighted average of key metrics to produce a reliability score. This score is derived from a linear utility function that integrates monitored metrics into an overarching reliability evaluation, consistent with methodologies in the literature \cite{utility_function}.

\paragraph{\textbf{Utility Function Definition}}
The reliability utility function, denoted as \(U_{\text{reliability}}\) is formulated at a specific time \(t\) and is expressed as:
    \begin{equation}
    \begin{split}
       U_{\text{reliability}}\left(\boldsymbol{\theta}(t) \mid \boldsymbol{\phi}\right) = \sum_{i=1}^{N} w_i \cdot u_i\left(\theta_i(t) \mid \boldsymbol{\phi}\right)
    \end{split}
    \end{equation}

where \(\boldsymbol{\theta}(t)\) represents the metrics vector at time \(t\), \(\boldsymbol{\phi}\) symbolizes additional parameters affecting the utility function, \(w_i\) corresponds to the weight of each metric, ensuring \(\sum_{i=1}^{N} w_i = 1\), and \(u_i(\theta_i(t) | \phi)\) denotes the individual utility functions for each metric. These utility functions capture reliability by translating monitored metrics into reliability rates, thus reflecting the probability of system stability at time \(t\). 

\paragraph{\textbf{Scoring Methodology}} \label{subsec: scoring}
Our scoring system involves continuous monitoring of three primary metrics: restart count, variability in response time, and variability in memory usage. The reliability scores derived from these metrics inform the dynamic adjustment of Pod replicas, enabling the system to respond promptly to workload fluctuations and reliability changes. The steps are as follows:

\begin{enumerate}
    \item \textbf{Metric Retrieval:} Metrics are collected using specific queries from the monitoring system (step 7 in Algorithm~\ref{alg:reliabilityScoringWithScaling}).
    \item \textbf{Metric Normalization:} The metrics are normalized linearly between 0 and 1, utilizing the utility function \(u\) specific to each metric \(m\) in the metric set \(I\). For every version \(v\) in the version set \(V\), the function is defined as:
    \begin{equation}
    \begin{split}
        \forall v \in V, \forall m \in I : u_{m}(v) &= 1 - \frac{\text{metric}_{m}(v) - \min(\text{metric}_{m}(V))}{\max(\text{metric}_{m}(V)) - \min(\text{metric}_{m}(V))}
    \end{split}
    \end{equation}

    \item \textbf{Reliability Score Calculation:} The reliability score for each version is computed (step 20 in Algorithm~\ref{alg:reliabilityScoringWithScaling}):

    \begin{equation}\label{eq:reliability_score}
    \begin{split}
        \text{Reliability\_Score}(v) &= \text{responseTimeWeight} \times u_{\text{responseTime}}(v) \\
        &\quad + \text{restartWeight} \times u_{\text{restarts}}(v) \\
        &\quad + \text{memoryWeight} \times u_{\text{memoryUsage}}(v)
    \end{split}
    \end{equation}
\end{enumerate}

\paragraph{\textbf{Replica Allocation Strategy}} \label{sec: scaling}

The system's architecture is designed to fine-tune the distribution of replicas among software versions in alignment with their reliability scores, maintaining at least one replica per version to prevent single-version failures. This is important as reliability can fluctuate over time, and maintaining at least one replica for each version safeguards against potential failures in other versions.
The replica allocation process, as detailed in \texttt{AdjustReplicaDistribution} function in Algorithm~\ref{alg:adjustReplicas}, involves a two-step adjustment: firstly, the proportional number of replicas for each software version is calculated based on its reliability score, and secondly, it fine-tunes this distribution to match the total number of replicas required.

\subsubsection{\textbf{Adaptive Scaling for Changing Workloads}} \label{sec:dynamic_workload}
Another factor that may impact the system's reliability is the changing workload.
Autoscaling is primarily done to maintain performance, but indirectly, if done properly, it will also improve system reliability. When scaling out/in, the main question would be which version should be scaled. Algorithm \ref{alg:reliabilityScoringWithScaling} details the logic of our scaling engine.

\begin{algorithm}
\caption{Version-Aware Autoscaling}\label{alg:reliabilityScoringWithScaling}
\begin{algorithmic}[1]
\small
\Require $MONITORING\_TIME$, $ACTION\_TIME$
$MAX\_REPLICAS$, $MIN\_REPLICAS$
$TOTAL\_REPLICAS$

\While{True}
\If{elapsed $=$ $MONITORING\_TIME$}
\State $currentCPU \gets getCPU()$
\State $currentScalingState \gets scalingAction(currentCPU)$
\State $scaleHistory$.add($currentScaleState$)

\For{deploymentVersion in deploymentVersions}
\State getPrometheusData(deploymentVersion)
\EndFor
\ElsIf{elapsed $=$ $ACTION\_TIME$}
\State $scaleAction \gets decideScaleBasedOnHistory(scaleHistory)$
\If{$scaleAction$ = Increase \textbf{and} \\
\hskip\algorithmicindent $TOTAL\_REPS < MAX\_REPS$}
\State $TOTAL\_REPLICAS$ =+ 1
\ElsIf{$scaleAction$ = Decrease \textbf{and} \\
\hskip\algorithmicindent $TOTAL\_REPS > MIN\_REPS$}
\State $TOTAL\_REPLICAS$ =- 1
\EndIf
\State $reliability\_scores \gets$ empty array
\For{$\text{deploymentVersion}$ in $\text{deploymentVersions}$}
\State Compute metrics \& reliabilityScore for deploymentVersion
\State $reliability\_scores$.add($reliabilityScore$)
\EndFor

\State \Call{AdjustReplicaDistribution}{deploymentVersions}
\EndIf
\EndWhile
\end{algorithmic}
\end{algorithm}

We use a threshold-based approach based on aggregate CPU utilization for dynamic scaling decisions (see \texttt{scalingAction} function in Algorithm~\ref{alg:scalingDecision}). To mitigate the ping-pong effect in scaling, our approach includes a historical analysis, as outlined in the \texttt{decideScaleBasedOnHistory} function in Algorithm~\ref{alg:scalingDecision}.

\begin{algorithm}
\caption{Selecting the Scaling Action}\label{alg:scalingDecision}
\begin{algorithmic}[1]
\small
\Require
\State $MAX\_CPU \gets Y\%$
\State $MIN\_CPU \gets X\%$
\Function{scalingAction}{$cpu$}
    \If{$cpu > MAX\_CPU$}
        \State \Return Increase
    \ElsIf{$cpu < MIN\_CPU$}
        \State \Return Decrease
    \Else
        \State \Return NoChange
    \EndIf
\EndFunction

\Function{decideScaleBasedOnHistory}{$history$}
    \State $increaseCount \gets 0$
    \State $decreaseCount \gets 0$
    
    \For{$action$ in $history$}
        \If{$action =$ Increase}
            \State $increaseCount$ =+ 1
        \ElsIf{$action =$ Decrease}
            \State $decreaseCount$ =+ 1
        \EndIf
    \EndFor
    
    \If{$decreaseCount > 2$}
        \State \Return Decrease
    \ElsIf{$increaseCount > 1$}
        \State \Return Increase
    \Else
        \State \Return NoChange
    \EndIf
\EndFunction
\end{algorithmic}
\end{algorithm}

\begin{algorithm}
\caption{Adjusting Replicas Based on Reliability Scores}\label{alg:adjustReplicas}
\begin{algorithmic}[1]
\Require $TOTAL\_REPLICAS$
\Function{AdjustReplicaDistribution}{deploymentVersions}
\State $total\_score \gets 0.0$
\For{score in reliability\_scores}
    \State $total\_score$ =+ $score$
\EndFor

\State newReplicas $\gets$ array of zeros for each deployment version
\State fractionalParts $\gets$ array of zeros for each deployment version
\State $allReplicas \gets 0$

\For{index, score in reliability\_scores}
    \State $proportionalReplica \gets \frac{TOTAL\_REPLICAS \times score}{total\_score}$
 
    \State newReplicas[index] $\gets$ floor of $proportionalReplica$
    \State fractionalParts[index] $\gets proportionalReplica - \text{newReplicas[index]}$

    \If{newReplicas[index] = 0}
        \State newReplicas[index] $\gets 1$
    \EndIf
    \State $allReplicas$ =+ $\text{newReplicas[index]}$
\EndFor

\State Sort indices of fractionalParts in descending order
\State $difference \gets allReplicas - TOTAL\_REPLICAS$

\If{$difference < 0$}
    \For{$i = 0$ to $-difference$}
        \State newReplicas[indices[i]] =+ 1
    \EndFor
\ElsIf{$difference > 0$}
    \For{$i = \text{len(indices)} - 1$ to $\text{len(indices)}-difference$}
        \If{newReplicas[indices[i]] > 1}
            \State newReplicas[indices[i]] =- 1
        \EndIf
    \EndFor
\EndIf
\EndFunction
\end{algorithmic}
\end{algorithm}

\subsection{Diversity Factor: Quantifying Version Variation}
In the adaptive scaling process, especially under variable workloads, a primary consideration is determining which software version to scale. A straightforward approach might be to favor scaling the most reliable versions. However, this strategy, while seemingly efficient, overlooks the inherent unpredictability of software behaviour. Sole reliance on a single version may expose the system to unforeseen vulnerabilities or issues specific to that iteration. In this context, the Diversity Factor (DF) emerges as an essential metric, serving as a measure of version diversity across our deployments. We define the DF as:
  \begin{equation}
    \begin{split}
        DF = \frac{1}{\sigma(R)}
    \end{split}
    \end{equation}
    
where \(\sigma(R)\) is the standard deviation of replica distribution. For instance, with versions \(V1\), \(V2\), and \(V3\) and replica counts \(R1\), \(R2\), and \(R3\):
  \begin{equation}
    \begin{split}
        \sigma(R) = \sqrt{\frac{(R1-\bar{R})^2 + (R2-\bar{R})^2 + (R3-\bar{R})^2}{3}}
    \end{split}
    \end{equation}

where \(\bar{R}\) is the average replica count.
DF's importance becomes evident in the context of the Algorithm \ref{alg:reliabilityScoringWithScaling}. When a scaling action is determined based on workload changes, the total number of replicas in the system is adjusted. However, instead of selectively increasing or decreasing specific versions, we apply the Algorithm \ref{alg:adjustReplicas}. This synergy between the two algorithms ensures that the distribution of these new total replicas takes into account both the reliability scores and the DF.

\section{Experimental Setup} \label{setup}
This section presents an in-depth overview of our experimental setup designed to evaluate the performance of our reliability engine and the version-aware autoscaling engine. First, we examine how our reliability engine maintains/improves the reliability of the software system under different types of chaos scenarios.
Then, we put our version-aware scaling engine under the test to show that it not only maintains the performance but also maintains/improves the reliability of software systems at different scales. More specifically, the scaling engine increases/decreases the population in such a way as to optimize the diversity factor. 

\subsection{Cluster Deployment and Configuration}
A Kubernetes cluster was deployed on two VMs in the Compute Canada cloud\footnote{\href{https://arbutus.cloud.computecanada.ca}{https://arbutus.cloud.computecanada.ca}}, consisting of one master and one worker node, each equipped with Ubuntu 22.04.2, 15 GB RAM, 4 VCPUs, and a total of 103 GB storage.

\subsubsection{\textbf{Initial Deployment and Configuration}}
To operationalize our system, a deployment named ``ReplicaBalancer'' is created as in Listing~\ref{lst: apply_deployment}. ``ReplicaBalancer'' acts as our reliability/scaling engine. The deployment is initiated with a Kubernetes command and configured with Listing~\ref{lst: set_env} environment variables.

\begin{lstlisting}[language=bash, caption={Initiating the "ReplicaBalancer" deployment}, label={lst: apply_deployment}]
$ kubectl apply -f AppDeploymentFile.yaml
\end{lstlisting}

Once ``ReplicaBalancer'' is up and running, the following command sets the environment variables that steer its behaviour.
\begin{lstlisting}[language=bash, caption={Setting environment variables for "ReplicaBalancer"}, label={lst: set_env}]
$ kubectl set env deployment/ReplicaBalancer \
DEPLOYMENT_IMAGES_REPLICAS=
"imageName1*replica1,imageName2*replica2,..." 
TOTAL_REPLICAS=9 
MONITORING_TIME=30s ACTION_TIME=2m 
MAX_REPLICAS=24 MIN_REPLICAS=3 
MAX_CPU=60 MIN_CPU=20 
SCALING=true
\end{lstlisting}

\subsubsection{\textbf{Parameterization Overview}} \label{sec: parameters}
The parameters employed in our experiments such as \texttt{TOTAL\_REPLICAS}, \texttt{MONITORING\_TIME}, and CPU thresholds, were selected to demonstrate the capabilities of our approach within the context of our experimental environment. It is important to note that the specific choice of these parameters was not the focal point of this research. These parameters can be fine-tuned according to the specific requirements of the user and the system.

Here's an overview of environment variables:
\begin{itemize}
    \item \texttt{DEPLOYMENT\_IMAGES\_REPLICAS}: Defines replica distribution for Docker images. If not specified, the system evenly allocates the total replicas among provided images.
    
    \item \texttt{MONITORING\_TIME}: Sets monitoring frequency.

    \item \texttt{ACTION\_TIME}: Specifies the interval for system response actions.
    
    \item \texttt{TOTAL\_REPLICAS}: Indicates initial replica count, adjustable based on workload.
    
    \item \texttt{MAX\_REPLICAS \& MIN\_REPLICAS}: Define the maximum and minimum replica limits.
    
    \item \texttt{MAX\_CPU \& MIN\_CPU}: Set CPU utilization thresholds for scaling.
    
    \item \texttt{SCALING}: Toggles autoscaling based on observed workload.
\end{itemize}

\subsection{Subject Systems}
\begin{figure}[htbp]
    \includegraphics[width=0.48\textwidth]{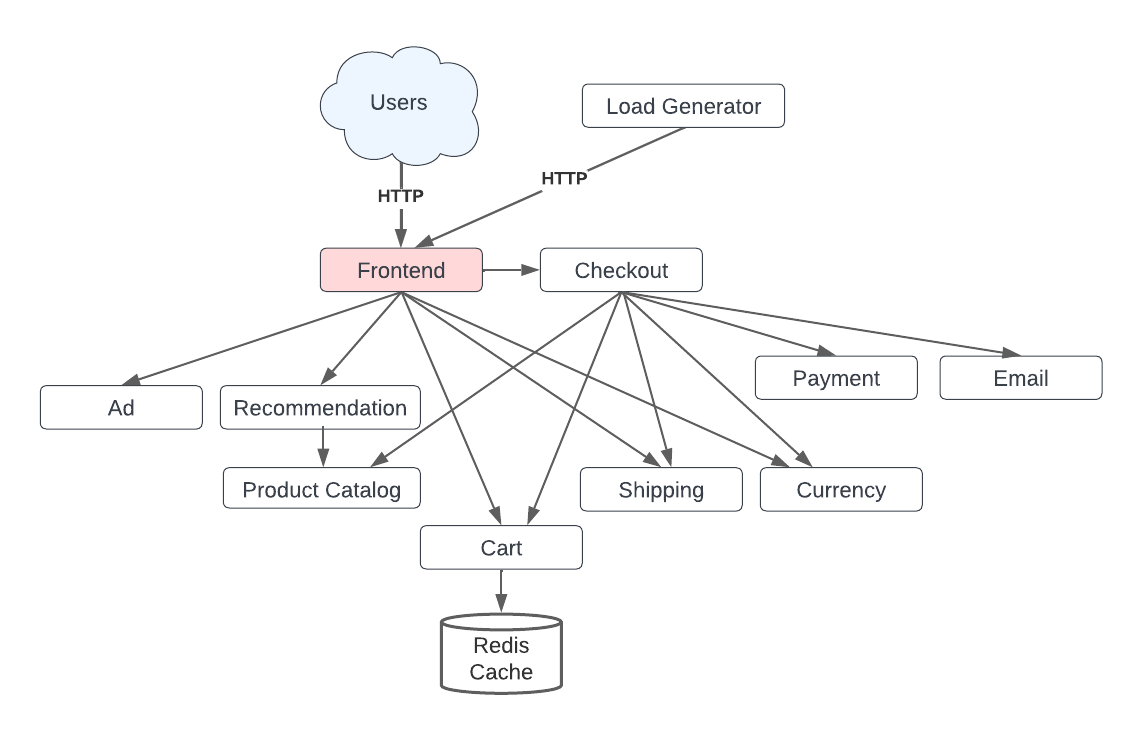}
    \caption{This visualization showcases the layout and interconnections of various microservices in the Online Boutique application.}
    \label{fig:online_boutique}
    \Description{This diagram visually represents the architecture of the Online Boutique, a microservices-based application, showcasing how various services are interconnected to handle e-commerce operations efficiently.}
\end{figure}

We analyzed the Online Boutique application\footnote{\href{https://github.com/GoogleCloudPlatform/microservices-demo}{https://github.com/GoogleCloudPlatform/microservices-demo}}, a cloud-native microservices application from Google. The overall architecture is depicted in Figure \ref{fig:online_boutique}.

\subsection{Workload}
We used Locust to simulate a typical e-commerce user flow. The workload consisted of 100 virtual users performing actions such as browsing products, adding items to the cart, and completing purchases. Requests were sent at random intervals to mimic real user behavior. During the experiments, we collected metrics such as response time, CPU usage, and memory consumption. The data was then analyzed to assess the performance and reliability of our proposed autoscaling approach.

\subsection{Critical Microservice Identification}
Choosing critical microservices is essential in multi-versioning, as it can impact resource allocation, budget and overall system reliability. Our study designates the frontend service as crucial because it is the primary user interface and function in aggregating data from various backend services. This decision aligns with PageRank-based methods that assess microservice importance (\cite{Liu1, inoue2005ranking, shi2020location}). While we do not delve into these methods, interested readers can consult the referenced studies. By concentrating on the frontend, we aim to deliver a robust user experience and maintain the operational integrity of the application on a manageable budget.

\subsection{Weight Configuration for Reliability Scoring} \label{sec: weights}
An integral part of our reliability scoring system is the assignment of specific weights to different metrics, which play a pivotal role in the overall reliability assessment. The weights assigned to each metric are as follows:

\begin{itemize}
    \item \textbf{restartWeight}: 0.5
    \item \textbf{memoryWeight}: 0.3
    \item \textbf{responseTimeWeight}: 0.2
\end{itemize}

The weights assigned in Equation \ref{eq:reliability_score} are based on the influence of each metric on the overall reliability of the Kubernetes system. Pod failure, weighted most heavily at 0.5, is crucial because it directly impacts system reliability. Frequent pod restarts can indicate underlying stability issues, leading to service disruption and reduced reliability \cite{li2021kubernetes}. The weight of 0.3 for memory usage variability addresses the importance of detecting memory leaks. Memory leaks can lead to system degradation or crashes, critically impacting reliability \cite{costa2023software}. Response time variability, with a lower weight of 0.2, affects user experience and system responsiveness. Although important, it's less directly related to the core operational reliability than pod failures and memory leak issues, hence has lower priority.

\subsection{Chaos Mesh Overview} \label{sec: ChaosMesh}
Chaos Mesh\footnote{\href{https://chaos-mesh.org}{https://chaos-mesh.org}} is an open-source platform for Chaos Engineering on Kubernetes. It simulates system faults and measures recovery processes, facilitated by a user-friendly web interface. The platform provides diverse chaos scenarios like network disruptions, system call failures, and resource constraints. Our experiments utilize Pod Chaos, HTTP Chaos, and Stress Chaos to emulate common failure modes in microservices.

\subsection{Chaos Experimentation} \label{sec: chaos_injection}

Given that different software versions may exhibit unique bugs affecting reliability, we simulate multi-versioning by injecting specific types of chaos into each version. This approach allows us to model distinct performance and reliability characteristics without the need to develop separate version functionalities. We align our chaos types with the key metrics defining reliability: restart counts, response time variability, and memory usage variability. Consequently, in our experimentation, three frontend microservice versions named ``Faulty'', ``InconsistentResponse'', and ``MemoryLeak'' are considered. We start the experiment with 3 healthy identical versions, and then, using Chaos Mesh, we introduce the above-mentioned bugs in the replicas to imitate the multi-versioning in software systems. Each version suggests the chaos linked to it:
\begin{enumerate}
\item \textbf{Faulty Version:} This version simulates the impact of bugs that cause frequent Pod restarts. Frequent restarts disrupt service continuity, leading to higher downtime and reduced reliability. This is implemented using \textbf{Pod Chaos} that targets the ``frontend-faulty-deployment'' every 3 minutes, terminating associated Pods for 30 seconds.

\item \textbf{InconsistentResponse Version:} Shows variable response times. Through \textbf{HTTP Chaos}, which introduces a 2-second delay in responses from the ``frontend-inconsistent-response-deployment'' every 4 minutes for 2 minutes, we demonstrate how variability in response times can reduce the predictability and performance standards expected from reliable systems. Such inconsistencies also impact user experience.

\item \textbf{MemoryLeak Version:} Faces a memory leak, pointing out its robustness limitations. This is explored using \textbf{Stress Chaos}, imposing memory stress on ``frontend-memory-leak-deployment'' every 4 minutes with two workers consuming 20MB memory each for 2 minutes. Memory leaks threaten the reliability of a system by compromising its ability to maintain operational performance over time.
\end{enumerate}

\section{Experimental Evaluation} \label{results}
To comprehend the system's behaviour, we conducted two experiments. The first experiment observed changes in replica distribution based on the reliability of individual components by introducing specific chaos. The second experiment examined the system's response to varying workloads, focusing on adaptability in replica scaling with changes in CPU utilization.

\subsection{Experiment 1: Evolution under Constant Workload}

\begin{figure}[htbp]
    \centering
    \includegraphics[width=0.47\textwidth]{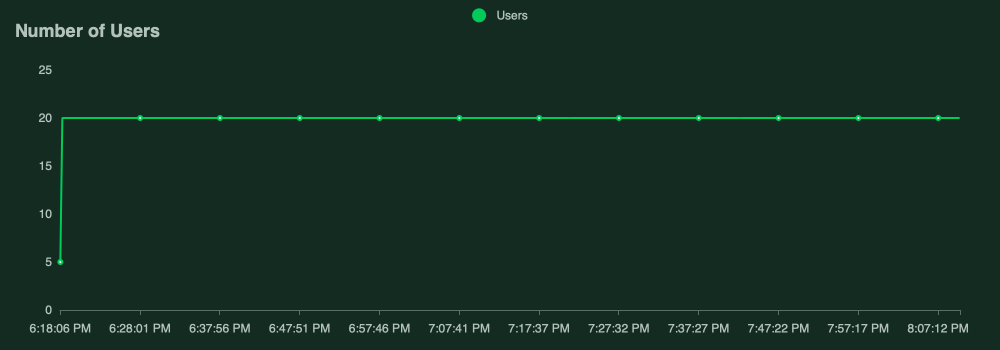}
    \caption{Number of users. This chart presents the number of users accessing the system during the first experiment.}
    \label{fig:number_of_users}
    \Description{This chart tracks user activity over time, providing a visual representation of the load on the system during the first experiment. It helps in analyzing how the system handles varying numbers of users.}
\end{figure}

In the first experiment, we studied the system under constant workload conditions with 20 concurrent users over 2 hours (Figure \ref{fig:number_of_users}).
For clarity and to observe the direct influence of each metric on reliability, we first injected only one type of chaos into each software version. In this way, we isolated the effects of each chaos type, making it easier to understand the specific impact of each metric on the system's overall reliability. After that, we injected all three types of chaos into the system and observed its behaviour again.

\begin{figure}[htbp]
    \centering
    \includegraphics[width=0.46\textwidth]{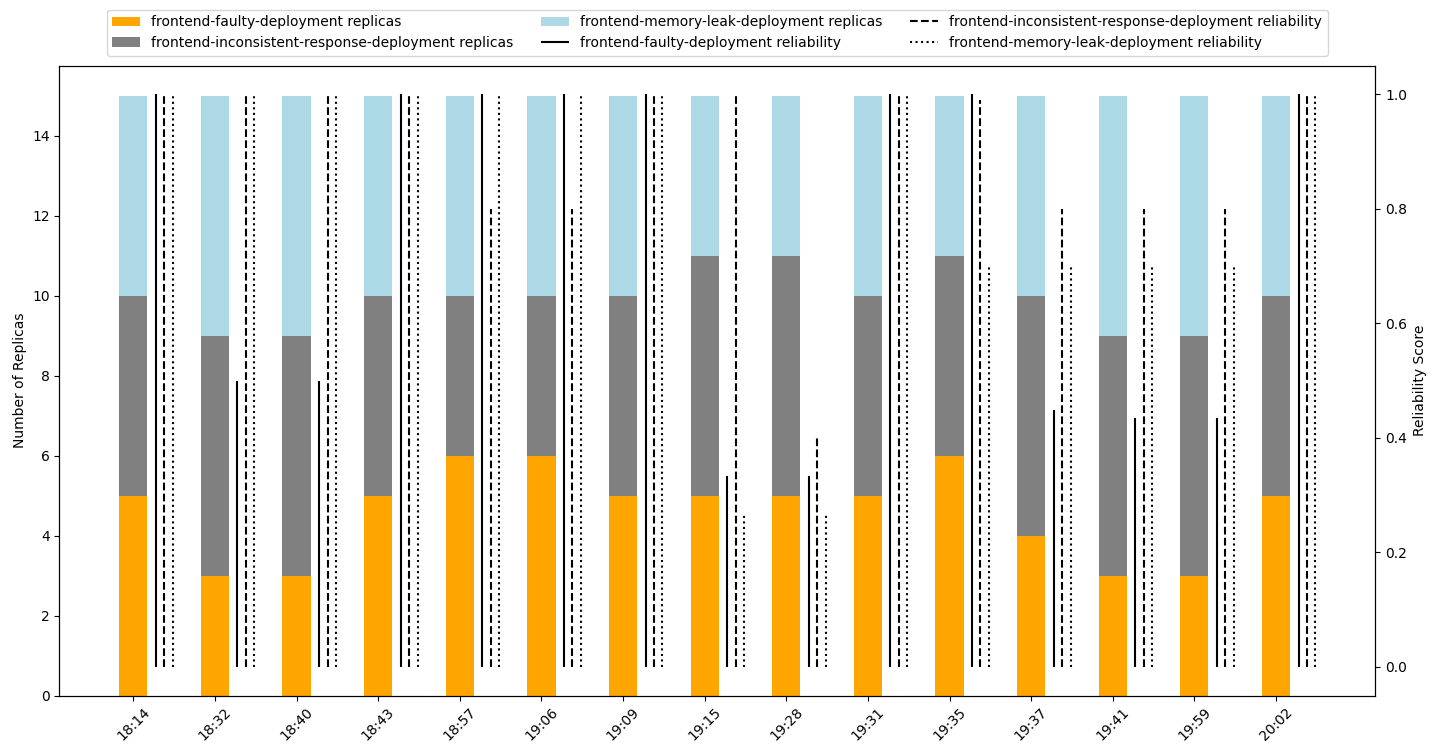}
    \caption{Replica and reliability over time. The bar chart illustrates the number of replicas for different frontend microservice versions over time. Adjacent vertical lines, differentiated by line style, represent the reliability scores for each version.}
    \label{fig:replicas}
    \Description{The bar chart illustrates the number of replicas for different frontend microservice versions over time, alongside their respective reliability scores, demonstrating the adaptive response of the system to varying conditions.}
\end{figure}

Initially, we allow the system to run undisturbed to observe the replica distribution across different versions. As shown in Figure \ref{fig:replicas}, each version started with 5 replicas, given a total of 15 replicas distributed equally. This uniform distribution can be attributed to the identical reliability scores of the versions, as no chaos had been introduced at this point and versions were identical.

\begin{figure}[htbp]
    \centering
    \includegraphics[width=0.47\textwidth]{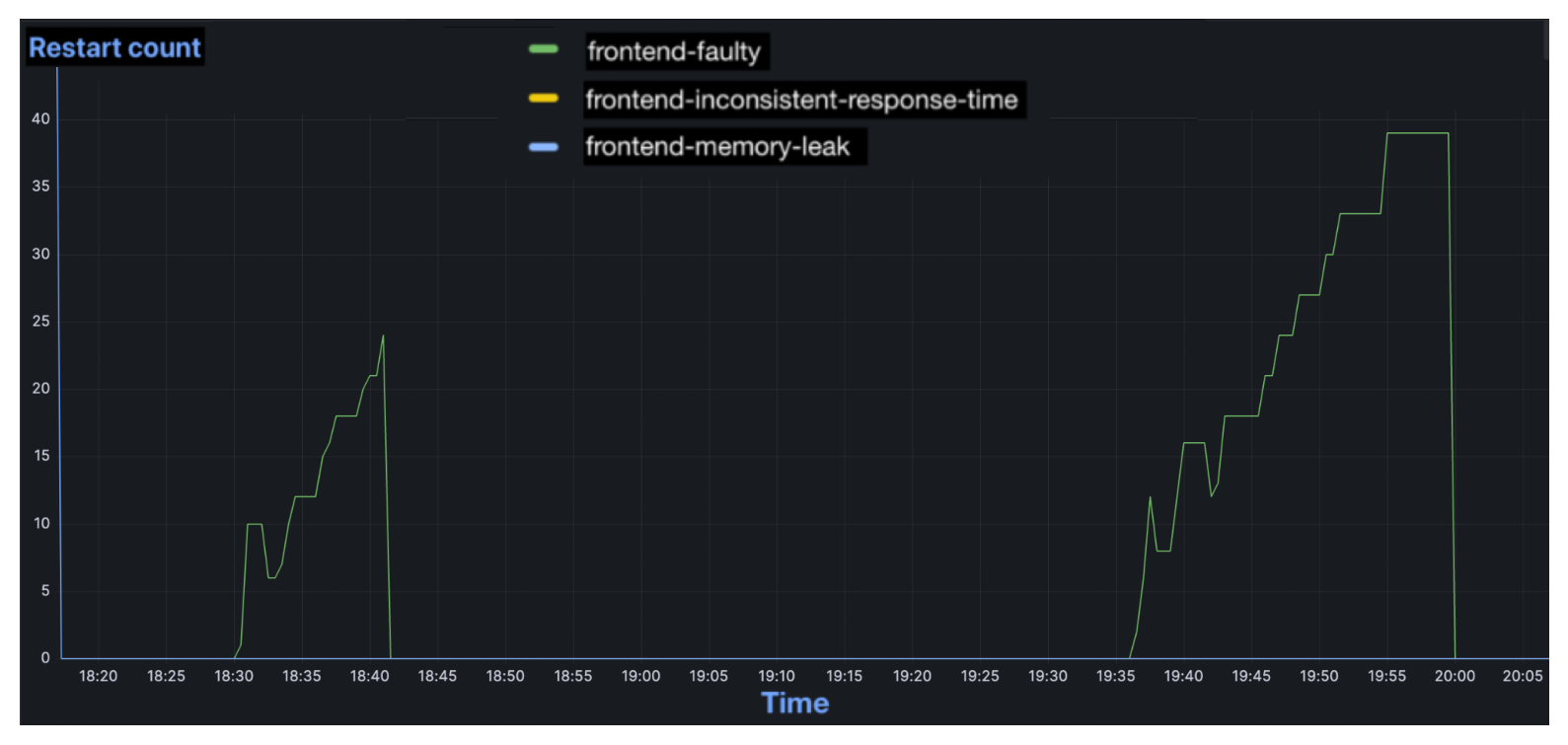}
    \caption{Restart count of frontend microservice versions. This chart illustrates the frequency and patterns of system restarts over a specific period.}
    \label{fig:restarts}
    \Description{This chart illustrates the restart frequency for different frontend microservice versions under chaos conditions, highlighting the system's resilience or vulnerability.}
\end{figure}

\begin{figure}[htbp]
    \centering
    \includegraphics[width=0.47\textwidth]{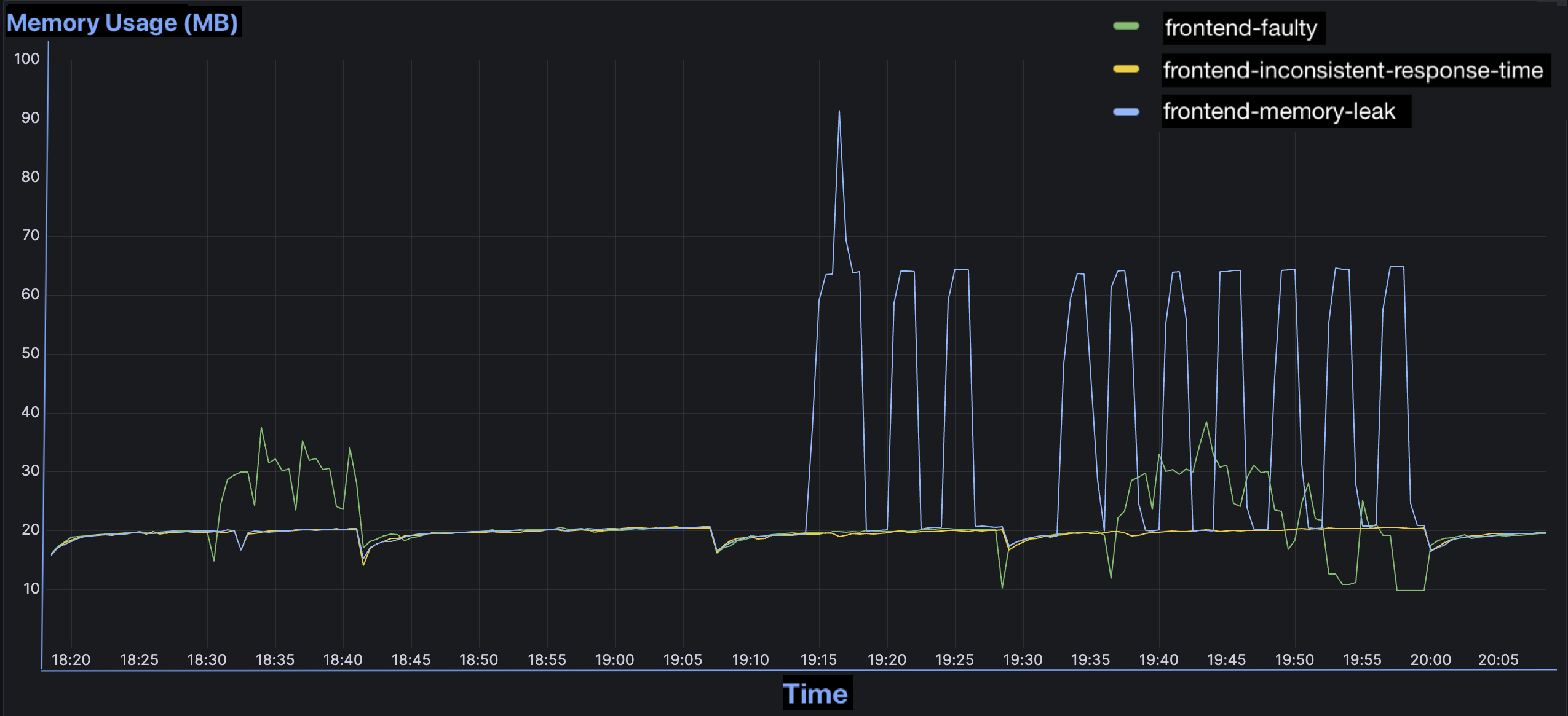}
    \caption{Memory usage over time (measured in MB). This graph provides a comprehensive look at the memory consumption patterns for different frontend deployments.}
    \label{fig:memory}
    \Description{This graph provides a comprehensive look at the memory consumption patterns for different frontend deployments.}
\end{figure}

Subsequently, we introduce Pod chaos, detailed in Section \ref{sec: chaos_injection}. Following this, a noticeable restart increase for the frontend-faulty-deployment was evident, as depicted in Figure \ref{fig:restarts}. Additionally, as seen in Figure \ref{fig:memory}, memory usage patterns for this version fluctuated. Simultaneously, Figure \ref{fig:replicas} captures the system's adaptive response regarding replica distribution. The faulty deployment's replica count was adjusted to 3. This change underscores the system's recognition of the diminished reliability of the faulty deployment due to the Pod chaos, which was an expected outcome. It's crucial to note that, at this juncture, only the Pod chaos was in play.

\begin{figure}[htbp]
    \centering
    \includegraphics[width=0.47\textwidth]{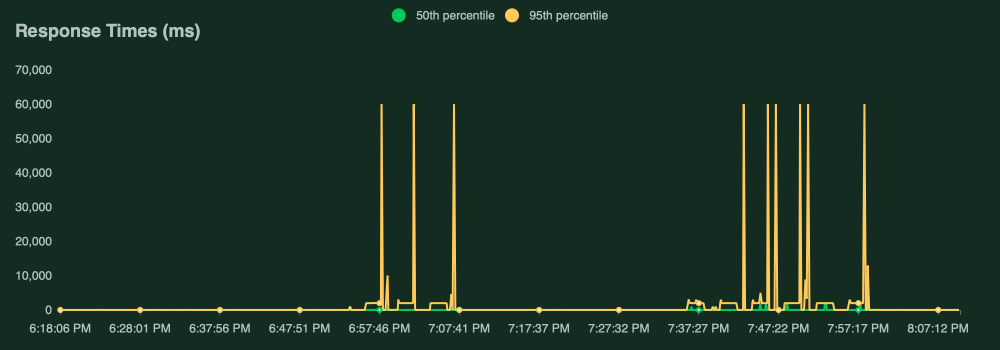}
    \caption{Application's response time during the first experiment.}
    \label{fig:response_times}
    \Description{This chart displays how quickly the system processes requests under varying conditions of chaos and load, providing insights into the impact of each chaos type on system responsiveness.}
\end{figure}

Stopping the chaos which replicating a scenario where a developer fixes a bug, the system gradually returned to an even distribution of 5 replicas per version.

Next, we applied HTTP chaos targeting the frontend-inconsistent-response-deployment. This resulted in noticeable system latency, as captured in Figure \ref{fig:response_times}. Interestingly, a replica was reallocated from frontend-inconsistent-response to frontend-faulty-deployment. Upon halting this chaos, the system returned to its balanced state of 5 replicas for each version.

In a subsequent test, we introduced stress chaos to the frontend-memory-leak-deployment. As seen in Figure \ref{fig:memory}, captured a rise in memory consumption. The result on replica distribution is illustrated in Figure \ref{fig:replicas}, which involves the subtraction of a replica from frontend-memory-leak and its addition to frontend-inconsistent-response. Following 16 minutes after stopping this chaos, the system restored its equilibrium of 5 replicas per version.

In our final testing phase, we combined all three chaos types to understand their compounded effect. The replica distribution settled at 3, 6, and 6, influenced by the metric weights detailed in Section \ref{sec: weights}. After stopping all chaos injections, the system eventually returned to its initial balanced state.

\subsection{Experiment 2: Evolutionary-Aware Auto-Scaling}

Our second experiment was designed to understand the system's dynamic scaling capabilities in relation to variable workloads, with a focus on CPU usage. We aimed to see how this adaptive scaling behaviour impacts software diversity, reliability, and performance.

\begin{figure}[htbp]
    \centering
    \includegraphics[width=0.46\textwidth]{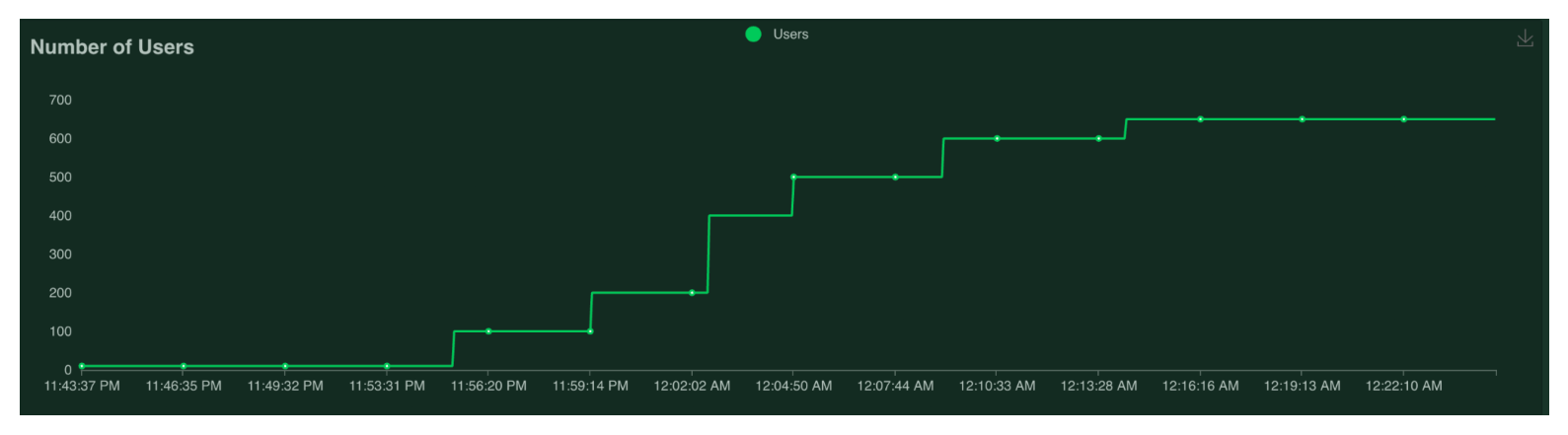}
    \caption{Number of users. This chart presents the number of active users accessing the system over a specific duration for the second experiment.}
    \label{fig:number_of_users2}
    \Description{This chart tracks the fluctuation of user activity throughout the experiment, correlating it with system load and scaling decisions.}
    
\end{figure}

\begin{figure}[htbp]
    \centering
    \includegraphics[width=0.46\textwidth]{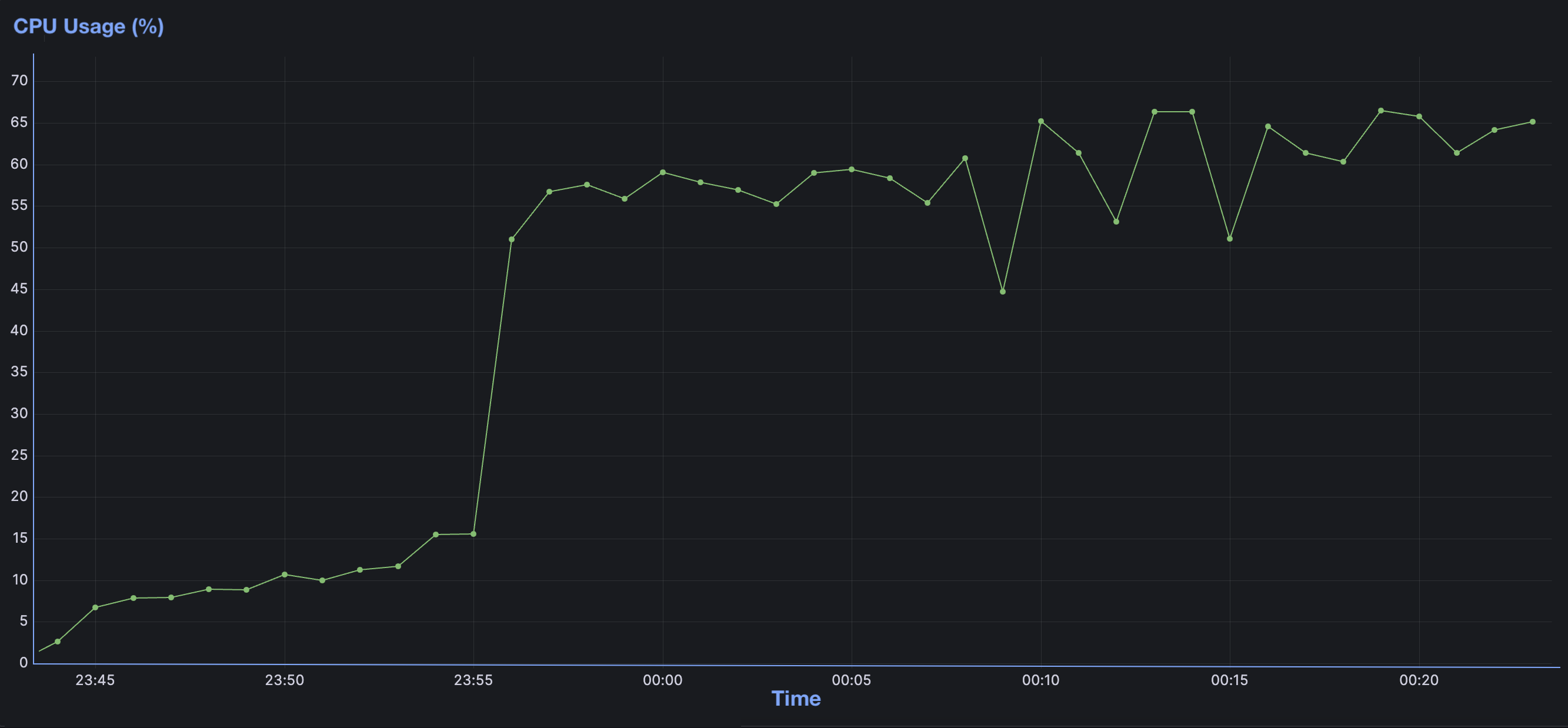}
    \caption{Average CPU utilization of frontend microservice Pods over time.}
    \label{fig:cpu}
    \Description{This graph depicts the CPU usage patterns, highlighting the correlation between user load and computational demand.}
\end{figure}

As the system was subjected to different user loads (Figure \ref{fig:number_of_users2}), we closely monitored the frontend microservice Pods' CPU usage as illustrated in Figure \ref{fig:cpu}. Over time, a direct correlation was observed between the workload and CPU usage as we changed the number of users. As the workload intensified, there was a consequent increase in CPU usage. In line with the user-configurable thresholds discussed in Section~\ref{sec: parameters}, our configuration set an upper CPU limit at 60\% and a lower limit at 20\%. Should the CPU usage exceed 60\%, the system would trigger a scale-out in the number of Pod replicas.

\begin{figure}[htbp]
    \centering
    \includegraphics[width=0.46\textwidth]{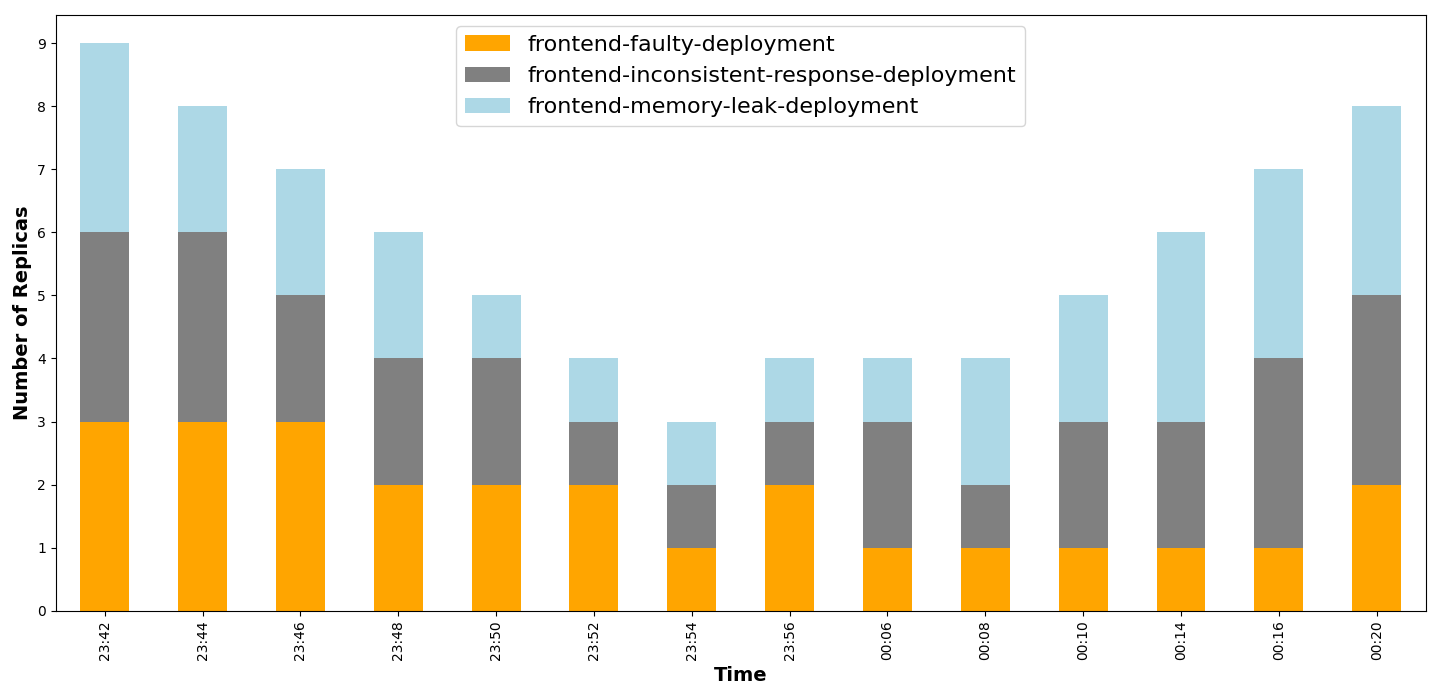}
    \caption{Dynamic scaling of frontend microservice Pods. This chart visualizes the system's dynamic scaling capabilities in response to varying workloads, highlighting the changes in the number of replicas over time. The system's adaptability to workload fluctuations is evident from the shifts in replica counts.}
    \label{fig:replicas2}
    \Description{This chart visualizes the system's dynamic scaling capabilities in response to varying workloads, highlighting the changes in the number of replicas over time. The system's adaptability to workload fluctuations is evident from the shifts in replica counts.}
\end{figure}

The early phase of the experiment, characterized by lower user numbers and CPU usage below the 20\% mark, saw a reduction in the total number of replicas, showcasing the system's efficiency and cost-effectiveness (Figure \ref{fig:replicas2}). As the experiment progressed with an incremental user load, the system dynamically scaled up, adding more replicas to handle the increased demand, thereby demonstrating its capability to maintain performance under varying workload conditions.

As seen in Figure~\ref{fig:replicas2}, during both scale-out and scale-in phases, the system's scaling decisions were informed by the reliability scores of different software versions, ensuring that replicas were allocated to preserve software diversity. Therefore, all three objectives, namely performance, cost, and reliability, were achieved simultaneously. 
 
\begin{figure}[htbp]
    \centering
    \includegraphics[width=0.46\textwidth]{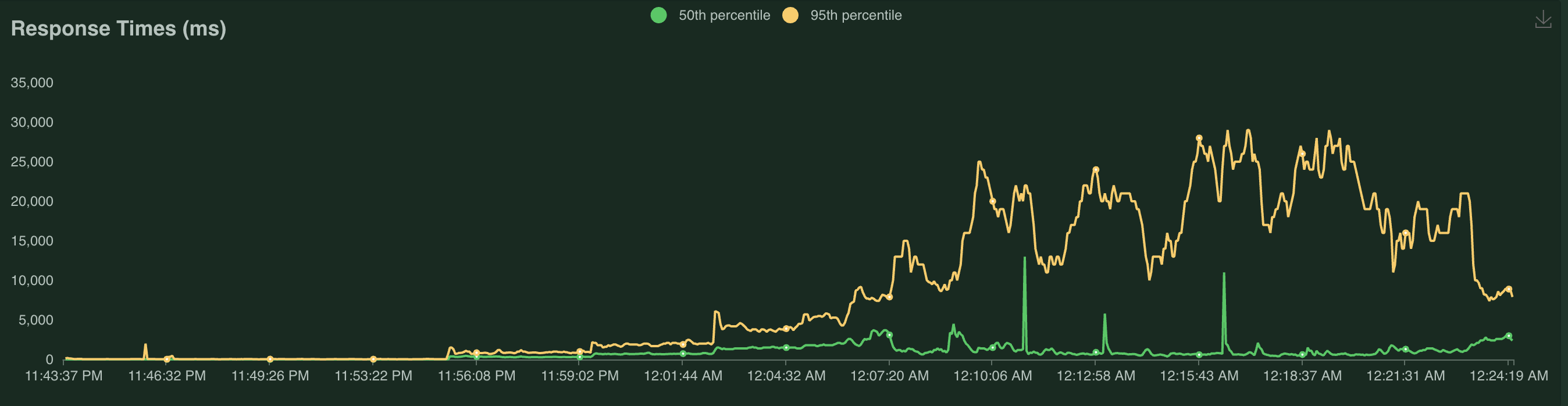}
    \caption{Application's response time during the second experiment.}
    \label{fig:response_times2}
    \Description{This graph illustrates the variation in response times as the system scales up or down, providing insights into the efficiency of the scaling algorithm under different loads.}
    
\end{figure}

\section{Related Work} \label{related}
In this section, we discuss prior work related to our research on software multi-versioning, containerized systems, microservices, reliability for microservices, and auto-scaling approaches.

\subsection{Software Multi-Versioning}

Software multi-versioning has been extensively researched. Larsen et al. \cite{b13} were among the pioneers, examining the effects of automated software redundancy on system security. This was followed by the work of Franz et al. \cite{b14}, who advocated for multi-versioning as a strategic defense against targeted cyber attacks. Persaud et al. \cite{b15} added a new dimension by integrating genetic algorithms with software redundancy for enhanced security. Cigsar et al. \cite{b16} explored multi-versioning to augment the reliability of repairable systems, while Gracie et al. \cite{b17} discussed its applications in ensuring system safety. Gorbenko et al. \cite{b18} utilized multi-versioning to improve the availability and reliability of web services. Borck et al. \cite{b19} introduced FEVIS, a program diversification method for detecting cyber attacks.

\subsection{Software Multi-Versioning for Containerized Systems}
In the realm of containerized systems, multi-versioning has been pivotal in enhancing system robustness. Wang et al. \cite{b20} advocated for its use in critical cloud software components to support fault tolerance, emphasizing cost-effective and selective application. Zheng et al. \cite{b22, b23} demonstrated the efficacy of multi-versioning in improving reliability and fault tolerance in service-oriented architectures. Gholami et al. \cite{b24} made a significant contribution with DockerMV, integrating multi-versioning into the Docker framework to facilitate application scaling. Dhaliwal and Khazaei \cite{10336238} focused on optimizing the performance of microservice systems by combining software multi-versioning with dynamic load balancing, demonstrating improvements in response times and resource utilization. The studies by Mohamed and El-Gayar \cite{mohamed2021end} and Pinto et al. \cite{pinto2018evaluating} further enriched this domain by assessing latency prediction and user acceptance in containerized environments.

\subsection{Microservices in Action}
The surge in microservices research has led to diverse applications. Timur et al. \cite{timur} proposed a distributed microservices-based solution for scalable deep learning facial recognition, leveraging Docker for data management. Lu et al. \cite{lu} developed a microservice platform for SSA data analytics, demonstrating its application through satellite-related tasks. Asaithambi et al. \cite{Asaithambi} introduced the Microservice-Oriented Big Data Architecture (MOBDA) for processing large-scale transportation data, focusing on Singapore's public transport system. Ali et al. \cite{Sajjad} explored modular microservices for real-time IoT-based health monitoring, exemplifying the adaptability of microservices in various sectors.

\subsection{Reliability for Microservices}
The aspect of microservice reliability has been a focal point in recent research endeavours. Vincenzo and Dragi \cite{Vincenzo} introduced a novel task unloading approach using Pareto optimization, targeting key performance metrics. Clab et al. \cite{clab} proposed a delay-adaptive strategy for replica synchronization, a critical factor in software reliability. Subsequent studies by Chen et al. \cite{chen2018uncertainty}, Liu et al. \cite{Liu1, Liu2}, and Pietrantuono et al. \cite{Pietrantuono} have significantly contributed to enhancing the reliability of microservice-based cloud applications, employing various methodologies including redundancy and circuit breakers. CoScal employs a reinforcement learning-based approach to optimize resource scaling for microservices, combining horizontal scaling, vertical scaling, and brownout techniques to handle complex workload scenarios effectively

\subsection{Auto-Scaling Approaches} \label{related: scaling-approaches}
The evolution of auto-scaling in cloud computing has seen a variety of innovative approaches. Al-Dhuraibi et al. \cite{b10} introduced ELASTICDOCKER, a reactive scaling strategy in Kubernetes, focusing on resource optimization and cost-effectiveness. Rodrigo et al. \cite{Calheiros} utilized the ARIMA model for accurate cloud workload prediction, while Messias et al. \cite{Messias} combined statistical methods with genetic algorithms for improved forecasting accuracy. Advanced predictive models like Bi-LSTM and AI-influenced techniques have been developed by researchers like Tang et al. \cite{Fisher}, Ming Yan et al. \cite{yan}, and Laszlo Toka et al. \cite{toka}. Dang-Quang and Yoo \cite{deeplearning, Multivariate-Yoo, Multivariate-Yoo2} and Dogani et al. \cite{dogani} have made notable contributions to proactive auto-scaling in Kubernetes, showcasing the potential of deep learning and attention-based models in this domain. Xu et al. \cite{9904920} proposed CoScal, a reinforcement learning-based approach that combines horizontal scaling, vertical scaling, and brownout techniques to handle complex workload scenarios effectively. While these auto-scaling mechanisms primarily focus on optimizing performance and cost, it is equally important to ensure that reliability is not compromised.

\section{Threats to Validity}\label{disc}
In this section, we discuss the potential threats to the validity of our experiment involving Chaos Mesh to test the reliability and robustness of our subject system.

\subsection{External Validity}

\textbf{Choice of Subject System:} Our experiments were conducted on the Online Boutique application within a specific system configuration. Future research should explore the generalizability of our findings across diverse setups and varying cluster sizes.

\noindent\textbf{Chaos Types Selection:} The chaos experiments were limited to certain types provided by Chaos Mesh. Real-world systems may encounter a broader and more complex range of disruptions not encompassed by our study.

\subsection{Internal Validity}

\textbf{Chaos Injection Timing:} The frequency of chaos injections in our tests may not mirror actual operational conditions, where failures could be more erratic or frequent.

\noindent\textbf{Replica Distribution:} We initiated our experiments with uniform replica distribution across software versions, which may not accurately reflect the varied distributions present in live environments.

\subsection{Construct Validity}

\textbf{Metric Selection:} We chose specific metrics to represent system reliability. While informative, these metrics may not translate universally to all systems, which could have different reliability benchmarks or operational criteria.

\noindent\textbf{Metric Weighting:} The weights given to each metric in Section \ref{subsec: scoring} are context-dependent. In different scenarios, the prioritization of these metrics could vary significantly.

\section{Future Work}\label{future}
\noindent \textbf{Extension to More Microservices:} Given the scale and complexity of software systems, future work could expand to include a wider range of microservices.	

\noindent \textbf{Refinement of Metrics and Validation:} While the current study relies on specific metrics like restart count, response time, and memory usage, there's room to investigate other metrics that might offer more comprehensive insights. This might also include a validation process for metrics selection across varied systems.

\noindent \textbf{Analysis of Real-world Traffic Patterns:} Incorporating actual user traffic patterns could provide a more accurate assessment of system behaviour under typical operational conditions.

\noindent \textbf{Dynamic Scaling Techniques:} There is an opportunity to improve upon the threshold-based scaling approach by integrating predictive models that enable anticipatory scaling actions.

\noindent \textbf{Multi-modal Metric Integration:} Future work could also include a mix of performance metrics, such as network bandwidth, disk I/O, and tailored application metrics for a comprehensive performance and scalability analysis.

\section{Conclusion} \label{conclusion}
Due to the substantial costs associated with implementing multiple system versions, multi-versioning has often been restricted to critical-mission systems.

However, microservice architecture allows for selective multi-versioning of critical components, making it viable across various software systems. This paper introduces a reliability engine designed to enhance system reliability through microservice version scores. This engine integrates seamlessly with any microservice-based application and incorporates a version-aware autoscaling mechanism that considers both performance and reliability. We validated our approach with practical cloud experiments. In this paper, we proposed a reliability engine that transparently maintains/improves the reliability of the system based on the reliability score of microservice versions. This engine can be augmented to any microservice-based application regardless of the internal logic of the underlying system. We also extend the reliability engine to conduct the scaling with the system's reliability in mind. Unlike conventional autoscaling, in which we are only concerned about performance, the Evolutionary-aware (i.e., version-aware) autoscaling engine proposed in this paper simultaneously satisfies performance and reliability at any scale. We conducted realistic experiments on the cloud to validate the proposed approach in this paper.

\bibliographystyle{IEEEtran}
\bibliography{ref}

\end{document}